\documentclass{ws-mpla}

\newcommand{\A}{{\mathcal{A}}}

\newcommand{\Od}{{\cal O}}

\newcommand{\lsim}   {\mathrel{\mathop{\kern 0pt \rlap
  {\raise.2ex\hbox{$<$}}}
  \lower.9ex\hbox{\kern-.190em $\sim$}}}
\newcommand{\gsim}   {\mathrel{\mathop{\kern 0pt \rlap
  {\raise.2ex\hbox{$>$}}}
  \lower.9ex\hbox{\kern-.190em $\sim$}}}

\begin{document}

\markboth{J. Beltr\'an Jim\'enez and A. L. Maroto}
{The Dark Magnetism of the Universe}

\catchline{}{}{}{}{}

\title{THE DARK MAGNETISM OF THE UNIVERSE}

\author{\footnotesize JOSE BELTR\'AN JIM\'ENEZ}

\address{D\'epartement de Physique Th\'eorique and Center for Astroparticle Physics,
Universit\'e de Gen\`eve\\ 24 quai Ansermet, CH--1211 Gen\`eve 4,
Switzerland.\\ Jose.Beltran@unige.ch }

\author{ANTONIO L. MAROTO}

\address{Departamento de F\'isica Te\'orica, Universidad Complutense de Madrid\\
28040, Madrid (Spain)\\
maroto@fis.ucm.es
}

\maketitle


\begin{abstract}
Despite the success of Maxwell's electromagnetism in the description of the electromagnetic
interactions on small scales, we know very little about the behaviour of electromagnetic fields
on cosmological distances. Thus, it has been suggested recently that the problems of dark energy 
and the origin of cosmic magnetic fields could
be pointing to a modification of Maxwell's theory on large scales. 
Here, we review such a proposal  in which 
the scalar state which is usually eliminated be means
of the Lorenz condition is allowed to propagate. On super-Hubble
scales,  the new mode is essentially given by the temporal component of the 
electromagnetic
potential  and contributes as an effective cosmological constant to the
energy-momentum tensor. The new state can be generated from
quantum fluctuations during inflation and it
is shown that the predicted value for the cosmological constant 
agrees with observations provided inflation took place at the
electroweak scale. We also consider more general theories including
non-minimal couplings to the space-time curvature in the presence
of the temporal electromagnetic background. 
We show that both in the minimal and non-minimal cases,  
the modified Maxwell's equations include  new
effective current terms which can generate magnetic fields 
from sub-galactic scales up to the present Hubble horizon.
The corresponding amplitudes could be enough to seed a galactic
dynamo or even to account for observations just by collapse and
differential rotation in the protogalactic cloud.

\end{abstract}


\section{Introduction}
Out of the four fundamental interactions, gravity and electromagnetism are the only long-range forces in nature. The standard theories for these two interactions, namely General Relativity (GR) and Maxwell's electromagnetism have received a vast experimental support in a huge range of scales. GR is able to explain the gravitational phenomena from sub-millimeters distances up to Solar System scales and standard electromagnetism has been probed from the tiny scales involved in high-energy colliders up to distances of order  1.3 A.U. corresponding to the coherence lengths of the magnetic fields dragged by the solar wind{\cite{limit}. However, despite the enormous success of these two theories, both of them present some unresolved problems when large scales are involved. For GR, cosmological observations require the presence of exotic components in the universe, i.e. dark matter and dark energy. Indeed, some attempts to account for such observations without invoking dark components include infrared modifications of GR. On the other hand,  the presence of cosmic magnetic fields found  in galaxies, clusters\cite{galactic1,galactic2,galactic3} 
and, 
very recently\cite{extragalactic1,extragalactic2,extragalactic3,extragalactic4}, also in the voids cannot be accommodated within Maxwell's theory. This could also be signaling that  a more careful analysis of the behavior
of electromagnetic fields in cosmological contexts is needed. 

A particularly interesting aspect is the quantization of gauge theories in non-trivial spacetimes. The usual quantization procedures of the electromagnetic field in flat spacetime rely on some type of subsidiary condition on the physical states to get rid of the unphysical gauge modes. Although all the approaches are equivalent in flat spacetime, such an equivalence has not been proven in curved background. In fact, the Gupta-Bleuler formalism for the covariant quantization has been shown to present difficulties in a time-dependent spacetime \cite{Parker,EM2}. The BRST method also exhibits similar pathologies in certain spacetimes\cite{ghosts1,ghosts2}. 

Here, we shall review a recent proposal of an extended theory of electromagnetism in which we can avoid the aforementioned difficulties. This extension is based on allowing the propagation of the state that is usually eliminated from the physical Hilbert space by means of the subsidiary condition. We shall show how this theory can be consistently quantized in an expanding universe with {\it three} physical states comprising the two polarizations of the usual photons plus one additional scalar state. The remarkable feature of the resulting theory is that the additional state can be generated from quantum fluctuations during inflation and give rise to and effective cosmological constant on large scales, whereas on sub-Hubble scales it leads to the generation of cosmic magnetic fields.

\section{Quantization in Minkowski spacetime}

The electromagnetic interaction is considered as a paradigm of a
well-behaved quantum field theory in Minkowski space-time.
Standard Maxwell's electromagnetism is the theory describing a pure
spin one massless particle. Nevertheless, the presence of the
local $U(1)$ symmetry in the kinetic term for the photon leads
to some difficulties when we want to quantize the theory. The
underlying reason for the appearance of these difficulties is the
redundancy in the description of the theory due to the $U(1)$
invariance so that we can be wrongly trying to quantize
superfluous degrees of freedom corresponding to pure gauge modes.
Let us briefly review these difficulties, since it will be useful
for our approach.

The starting point will be the Maxwell's action:
\begin{eqnarray}
S=\int d^4x \left(-\frac{1}{4}F_{\mu\nu}F^{\mu\nu}+ A_\mu
J^\mu\right) \label{Maxaction}
\end{eqnarray}
where $F_{\mu\nu}=\partial_\mu A_\nu-\partial_\nu A_\mu$ and $J_\mu$
is a conserved current. This action naturally arises when imposing the $U(1)$ gauge
symmetry in the sector of charged particles, leading to the symmetry $A_\mu\rightarrow A_\mu+\partial_\mu \chi$ with
$\chi$ an arbitrary function of space-time coordinates for the electromagnetic potential. The classical equations of motion derived from this action are the usual Maxwell's equations:
\begin{eqnarray}
\partial_\nu F^{\mu\nu}=J^\mu.
\label{Maxwell}
\end{eqnarray}
The quantization of this theory is not straightforward for a series of reasons that are connected, namely: the vanishing of the conjugate momentum of the temporal component makes not possible to write down its corresponding commutation relation and it is not possible to construct a propagator for the $A_\mu$ field.  As commented above, the underlying reason for  these problems is the presence of the gauge symmetry because that implies the presence of unphysical degrees of freedom which we have to get rid of before proceeding to the quantizations of the physical ones. In order to achieve that, two approaches are commonly used. On one hand, we have the gauge-fixing procedure where we use the gauge symmetry to impose a condition 
on the electromagnetic potential. Thus for example, we can impose the so-called Lorenz condition $\partial_\mu A^\mu=0$ so that the equations of motion reduce to:
\begin{eqnarray}
\Box A_\mu=J_\mu.\label{eqLo}
\end{eqnarray}
Since the Lorenz condition still leaves a residual gauge invariance
$A_\mu\rightarrow A_\mu+\partial_\mu \theta$, provided $\Box
\theta=0$, we can eliminate one additional component of the $A_\mu$ field in the
asymptotically free regions (typically $A_0$) which means $\vec
\nabla \cdot \vec A=0$. Thus the temporal and
longitudinal photons are removed and we are left with the two
transverse polarizations of the massless free photon, which are
the only modes (with positive energies) which are quantized in
this formalism.

On the other hand, in the covariant (Gupta-Bleuler) quantization approach, we start with a modified version of Maxwell's action, namely:
\begin{eqnarray}
S=\int d^4x \left(-\frac{1}{4}F_{\mu\nu}F^{\mu\nu}+\frac{\xi}{2}
(\partial_\mu A^\mu)^2+ A_\mu J^\mu\right).
 \label{actionGB}
\end{eqnarray}
that only preserves the residual gauge invariance. The equations of
motion obtained from this action now read:
\begin{eqnarray}
\partial_\nu F^{\mu\nu}&+&\xi\partial^\mu(\partial_\nu
A^\nu)=J^\mu. \label{fieldeq}
\end{eqnarray}
In order to recover Maxwell's equations, the Lorenz condition must be
imposed so that the $\xi$ term does not contribute. At the classical level
this can be achieved by means of appropriate boundary conditions
on the field because $\partial_\nu A^\nu$ evolves as a free scalar field, as obtained from the divergence of (\ref{fieldeq}):
\begin{eqnarray}
\Box(\partial_\nu A^\nu)=0
\end{eqnarray}
where we have made use of current conservation. Thus, if $\partial_\nu A^\nu$ vanishes for large $\vert t \vert$, it will vanish at
all times. At the quantum level, the Lorenz condition cannot be
imposed as an operator identity, but only in the weak sense
$\partial_\nu A^{\nu \,(+)}\vert \phi\rangle=0$, where $(+)$
denotes the positive frequency part of the operator and $\vert
\phi\rangle$ stands for a physical state. This condition is
equivalent to imposing $[{\bf a}_0(\vec k) +{\bf a}_\parallel(\vec
k)] |\phi\rangle=0$, with ${\bf a}_0$ and ${\bf a}_\parallel$ the
annihilation operators corresponding to temporal and longitudinal
electromagnetic states. Thus, in the covariant formalism, the
physical states contain the same number of temporal and
longitudinal photons, so that their energy densities, having
opposite signs, cancel each other. Thus, we see that also in this
case, the Lorenz condition seems to be essential in order to
recover standard Maxwell's equations and get rid of the negative
energy states. Now let us see what happens when moving on to an
expanding universe.




\section{Quantization in an expanding universe}
We shall consider a curved background and, in particular, 
an expanding universe, and we shall show how consistently imposing the Lorenz
condition in the covariant formalism turns out to be difficult to
realize \cite{Parker,EM2}. Let us consider the curved
space-time version of action (\ref{actionGB}):
\begin{eqnarray}
S=\int d^4x
\sqrt{g}\left[-\frac{1}{4}F_{\mu\nu}F^{\mu\nu}+\frac{\xi}{2}
(\nabla_\mu A^\mu)^2+ A_\mu J^\mu\right]
 \label{actionF}
\end{eqnarray}
leading to the corresponding equations of motion:
\begin{eqnarray}
\nabla_\nu F^{\mu\nu}+\xi\nabla^\mu(\nabla_\nu A^\nu)=J^\mu
\label{EMeqexp}
\end{eqnarray}
whose divergence imposes:
\begin{eqnarray}
\Box(\nabla_\nu A^\nu)=0.\label{minimal}
\end{eqnarray}
We see once again that $\nabla_\nu A^\nu$  behaves as a free scalar
field which is decoupled from the conserved electromagnetic
currents, but it is non-conformally coupled to gravity so that it can be excited
from quantum vacuum fluctuations by the expanding background in a
completely analogous way to the inflaton fluctuations during
inflation. Thus, this poses the question of the validity of the
Lorenz condition at all times.

In order to illustrate this effect, we will present a toy example.
Let us consider quantization in the absence of currents, in a
spatially flat expanding background, whose metric is written in
conformal time as $ds^2=a(\eta)^2(d\eta^2-d\vec x^2)$ with
$a(\eta)=2+\tanh(\eta/\eta_0)$ where $\eta_0$ is constant. This
metric has two asymptotically Minkowskian regions in the remote
past and far future. We  solve the coupled system of equations
(\ref{EMeqexp}) for the corresponding Fourier modes, which are
defined as $\A_\mu(\eta,\vec x)= \int
d^3k\A_{\mu \vec k}(\eta) e^{i\vec k \vec x}$. Thus, for a given
mode $\vec k$, the $\A_\mu$ field is  decomposed into temporal,
longitudinal and transverse components. The corresponding
equations read:
\begin{eqnarray}
\A_{0k}''-\left[\frac{k^2}{\xi}-2{\mathcal{H}}'
+4{\mathcal{H}}^2\right]\A_{0k}
-2ik\left[\frac{1+\xi}{2\xi}\A_{\parallel k}'
-{\mathcal{H}}\A_{\parallel k}\right]&=&0 \label{modes}\nonumber\\
\A_{\parallel k}''-k^2\xi\A_{\parallel
k}-2ik\xi\left[\frac{1+\xi}{2\xi}\A_{0k}'
+{\mathcal{H}}\A_{0k}\right]&=&0\nonumber\\
\vec{\A}_{\perp k}''+k^2\vec{\A}_{\perp k }&=&0
\end{eqnarray}
with ${\cal H}=a'/a$ and $k=\vert \vec k\vert$. We see that the
transverse modes are decoupled from the background, whereas the
temporal and longitudinal ones are non-trivially coupled to each
other and to gravity. Let us prepare our system  in an initial
state $\vert \phi\rangle$ belonging to the physical Hilbert space,
i.e. satisfying $\partial_\nu \A^{\nu \,(+)}_{in}\vert
\phi\rangle=0$ in the initial flat region. Because of the
expansion of the universe, the positive frequency modes in the
$in$ region with a given temporal or longitudinal polarization
$\lambda$ will become a linear superposition of positive and
negative frequency modes in the $out$ region and with different
polarizations $\lambda'$ (we will work in the Feynman gauge
$\xi=-1$). Thus, we have:
\begin{eqnarray}
\A_{\mu \vec k}^{\lambda \; (in)}=\sum_{\lambda'=0,\parallel}
\left[\alpha_{\lambda\lambda'}(\vec k) \A_{\mu \,\vec k}^{\lambda'
\; (out)}+\beta_{\lambda\lambda'}(\vec k) \overline{\A_{\mu
\,-\vec k}^{\lambda' \; (out)}}\,\right]
\end{eqnarray}
or in terms of creation and annihilation operators:
\begin{equation}
{\bf a}_\lambda^{(out)}(\vec k)=
\sum_{\lambda'=0,\parallel}\left[\alpha_{\lambda\lambda'}(\vec k)
{\bf a}_{\lambda'}^{(in)}(\vec
k)+\overline{\beta_{\lambda\lambda'}(\vec k)} {\bf
a}_{\lambda'}^{(in)\dagger}(-\vec k)\right]
\end{equation}
with $\lambda, \lambda'=0,\parallel$ and where
$\alpha_{\lambda\lambda'}$ and $\beta_{\lambda\lambda'}$ are the
so-called Bogolyubov coefficients (see Ref. \refcite{Birrell} for a
detailed discussion), which are  normalized in our case according
to:
\begin{eqnarray}
\sum_{\rho,\rho'=0,\parallel}(\alpha_{\lambda \rho}\,
\overline{\alpha_{\lambda'\rho'}}\, \eta_{\rho \rho'}
-\beta_{\lambda
\rho}\,\overline{\beta_{\lambda'\rho'}}\,\eta_{\rho\rho'})=\eta_{\lambda\lambda'}
\end{eqnarray}
with $\eta_{\lambda\lambda'}=diag(-1,1)$ with
$\lambda,\lambda'=0,\parallel$. Notice that the normalization is
different from the standard one \cite{Birrell}, because of the
presence of  negative norm states.

Thus, the system will end up in a final state which no longer
satisfies the weak Lorenz condition i.e. in the {\it out} region
$\partial_\nu \A^{\nu \,(+)}_{out}\vert \phi\rangle\neq 0$. This
is shown in Fig. \ref{photonsoutfig}, where we have computed  the
final number of temporal and longitudinal photons
$n_\lambda^{out}(k)=\sum_{\lambda'}\vert\beta_{\lambda\lambda'}(\vec
k) \vert^2 $, starting from an initial vacuum state with
$n_0^{in}(k)=n_\parallel^{in}(k)=0$. We see that, as commented
above, in the final region $n_0^{out}(k)\neq n_\parallel^{out}(k)$
and the state no longer satisfies the Lorenz condition. Notice
that the failure comes essentially from large scales ($k\eta_0\ll
1$), since on small scales ($k\eta_0\gg 1$), the Lorenz condition
can be restored. This can be easily  interpreted from the fact
that on small scales the geometry can be considered as essentially
Minkowskian.

\begin{figure}[ht!]
\begin{center}
{\epsfxsize=8cm\epsfbox{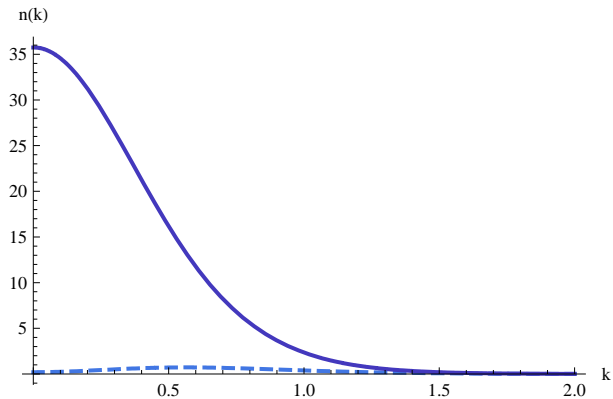}} \caption{\small Occupation
numbers for temporal (continuous line) and longitudinal (dashed
line) photons in the $out$ region vs. $k$ in $\eta_0^{-1}$ units.}\label{photonsoutfig}
\end{center}
\end{figure}

In order to overcome this problem, it would be possible to use the so 
called Faddeev-Popov quantization method \cite{Adler:1976jx,Brown:1986tj}. 
In this approach additional unphysical modes
(ghosts) are introduced in the theory. They are scalar fields but satisfying 
fermionic statistics. The physical states of the theory then  are
chosen as those for which the ghost contribution exactly cancels
the contribution from the unphysical gauge modes. However, it has 
been recently shown that in certain geometries, this type of cancellation
can be also problematic \cite{ghosts1,ghosts2}.

In the following, we shall follow a different approach in order to
deal with the difficulties found in the Gupta-Bleuler formalism
and  we shall explore the possibility of quantizing
electromagnetism in an expanding universe without imposing the
subsidiary Lorenz condition.

\section{Quantization without the Lorenz condition}
As we have shown in the previous section, consistently imposing the Lorenz condition is not an easy task. This could actually be signaling a subtle difficulty in formulating a gauge invariant theory of electromagnetism in non-trivial background spacetimes. Here we shall explore the possibility that the fundamental theory of
electromagnetism is not given by the gauge invariant
 action (\ref{Maxaction}), but by the gauge non-invariant action:
\begin{eqnarray}
S=\int d^4x
\sqrt{g}\left[-\frac{1}{4}F_{\mu\nu}F^{\mu\nu}+\frac{\xi}{2}
(\nabla_\mu A^\mu)^2+ A_\mu J^\mu\right].
\end{eqnarray}
Notice that although this action is not invariant under general
gauge transformations, it respects the invariance under residual
ones. Moreover, the dynamics of the ordinary transverse photons is not
affected, since the extra term only involves temporal and longitudinal polarizations. 
Let us emphasize that we
are assuming that the inclusion of the gauge-breaking term is not a
mathematical trick in order to quantize an otherwise
gauge-invariant theory, but that such a term is an essential part
of a gauge non-invariant electromagnetic theory. Since  the
fundamental electromagnetic theory is assumed non-invariant under
arbitrary gauge transformations, then there is no need to impose
the Lorenz constraint in the quantization procedure. Therefore,
having removed one constraint, the theory  contains one additional
degree of freedom.  Thus, the general solution for the modified equations (\ref{EMeqexp})
can be written as:
\begin{eqnarray}
\A_\mu=\A_\mu^{(1)}+\A_\mu^{ (2)}+\A_\mu^{(s)}+\partial_\mu \theta
\end{eqnarray}
where $\A_\mu^{(i)}$ with $i=1,2$ are the two transverse modes of
the massless photon, $\A_\mu^{(s)}$ is a new scalar state\footnote{Here we mean scalar in the sense that it is the only mode contributing to the divergence of the field so that $\nabla^\mu \A_\mu=\nabla^\mu \A_\mu^{(s)}$.} that
represents the mode that would have been eliminated if we had
imposed the Lorenz condition and, finally, $\partial_\mu \theta$
is a purely residual gauge mode, which can be eliminated by means
of a residual gauge transformation in the asymptotically free
regions, in a completely analogous way to the elimination of the
$A_0$ component in the Lorenz gauge.

In order to define the quantum theory, apart from the dynamics 
given by the above Lagrangian, we have to specify the 
Hilbert space of the physical states. Thus we follow a similar approach to the 
Lorenz or Coulomb gauge quantization procedure in which the gauge is fixed and then 
only the physical modes are quantized. In the present case, as commented before,  
the free theory contains 
three physical states and we perform the mode
expansion of the field with the corresponding creation and
annihilation operators:
\begin{eqnarray}
\A_{\mu}=\int d^3\vec{k}\ \sum_{\lambda=1, 2,s}\left[{\bf
a}_\lambda(k)\A_{\mu k}^{(\lambda)} +{\bf
a}_\lambda^\dagger(k)\overline{\A_{\mu k}^{(\lambda})}\, \right]
\end{eqnarray}
where the modes are required to be orthonormal with respect to the
scalar product\cite{Pfenning:2001wx}
\begin{eqnarray}
\left(\A^{(\lambda)}_k,\A^{(\lambda')}_{k'}\right)&=&i\int_{\Sigma}d\Sigma_\mu\left[\,
\overline{\A_{\nu k}^{(\lambda) }}\;\Pi^{(\lambda')\mu\nu}_{k'}-
\overline{\Pi^{(\lambda)\mu\nu }_{k}}\;\A_{\nu k'}^{(\lambda')}\right]\nonumber\\
&=&\delta_{\lambda\lambda'}\delta^{(3)}(\vec k-\vec k'),\;\;\;\;\;
\lambda,\lambda'=1,2,s\label{scalarproduct}
\end{eqnarray}
where $d\Sigma_\mu$ is the three-volume element of the Cauchy
hypersurfaces. In a Robertson-Walker metric in conformal time, it
reads $d\Sigma_\mu= a^4(\eta)(d^3x,0,0,0)$. The generalized
conjugate momenta are defined as:
\begin{eqnarray}
\Pi^{\mu\nu}=-(F^{\mu\nu}-\xi g^{\mu\nu}\nabla_\rho A^\rho)
\end{eqnarray}
Notice that the three modes can be chosen to have positive
normalization. The equal-time commutation relations:
\begin{eqnarray}
\left[\A_\mu(\eta,\vec x),\A_\nu(\eta,\vec x\,')\right]=
\left[\Pi^{0\mu}(\eta,\vec x),\Pi^{0\nu}(\eta,\vec x')\right]=0
\end{eqnarray}
and
\begin{eqnarray}
\left[\A_\mu(\eta,\vec x),\Pi^{0\nu}(\eta,\vec x\,')\right]=
i\frac{\delta_\mu^{\;\nu}}{\sqrt{g}}\delta^{(3)}(\vec x-\vec x\,')
\end{eqnarray}
can be seen to imply the canonical commutation relations
\begin{eqnarray}
\left[{\bf a}_\lambda(\vec{k}),{\bf
a}_{\lambda'}^\dagger(\vec{k'})\right]
=\delta_{\lambda\lambda'}\delta^{(3)}(\vec{k}-\vec{k'}),\;\;\;
\lambda,\lambda'=1,2,s
\end{eqnarray}
by means of the normalization condition in (\ref{scalarproduct}).  Notice that the sign of the commutators is positive for the three
physical states, i.e. there are no negative norm states in the
theory, which in turn guarantees that there are no negative energy
states as we will see below in an explicit example.

Notice that unlike the quantization in the Gupta-Bleuler formalism, in this
extended theory it is not possible to eliminate the unphysical degree of freedom in 
a manifestly covariant way.

Since  $\nabla_\mu\A^{\mu}$ evolves as a minimally coupled scalar
field, as shown in (\ref{minimal}), on sub-Hubble scales ($\vert
k\eta\vert \gg 1$), we find that for arbitrary background
evolution, $\vert \nabla_\mu\A^{(s)\mu}_k\vert \propto e^{\pm ik\eta} a^{-1}$,
i.e. the mode behaves as a plane-wave with decaying amplitude giving rise 
through  (\ref{EMeqexp}) to a longitudinal electric wave. This is an important 
difference with respect to ordinary electromagnetism, since the modified theory
allows the propagation of longitudinal electric waves in the absence of sources. 
As we will show below, these longitudinal fields in the presence of a highly conductive
plasma could induce the production of large scale magnetic fields.

On the other hand, on super-Hubble scales ($\vert k\eta\vert \ll
1$), $\vert\nabla_\mu\A^{(s)\mu}_k\vert= const.$ which implies
that the field contributes as a cosmological constant in
(\ref{actionF}). Indeed, the energy-momentum tensor derived from
(\ref{actionF}) reads:
\begin{eqnarray}
T_{\mu\nu}&=&-F_{\mu\alpha}F_\nu^{\;\;\alpha}
+\frac{1}{4}g_{\mu\nu}F_{\alpha\beta}F^{\alpha\beta}\nonumber\\
&+&\frac{\xi}{2}\left[g_{\mu\nu}\left[\left(\nabla_\alpha
A^\alpha\right)^2 +2A^\alpha\nabla_\alpha\left(\nabla_\beta
A^\beta\right)\right] -4A_{(\mu}\nabla_{\nu)}\left(\nabla_\alpha
A^\alpha\right)\right]
\end{eqnarray}
Notice that for the scalar electromagnetic mode in the
super-Hubble limit, the contributions involving $F_{\mu\nu}$
vanish and only the piece proportional to $\xi$ is relevant. Thus,
it can be easily seen that, since in this case $\nabla_\alpha
A^\alpha=constant$, the energy-momentum tensor is just given by:
\begin{eqnarray}
T_{\mu\nu}=\frac{\xi}{2} g_{\mu\nu}(\nabla_\alpha A^\alpha)^2
\label{Lambdaconf}
\end{eqnarray}
which is the energy-momentum tensor of a cosmological constant and
whose value is given by the four-divergence of the electromagnetic
field. In fact,  this result is not specific of FLRW spacetimes, but it is valid for any geometry. In a field configuration with the vector potential given by the gradient of a scalar that is not a pure residual gauge mode, i.e., $A_\mu=\partial_\mu\phi$ with $\Box\phi\neq0$, we obtain that $F_{\mu\nu}$ vanishes. Thus, from the equations of motion (\ref{EMeqexp}) in the absence of external currents we get that $\nabla_\mu A^\mu$ is constant so that the energy momentum tensor reduces once again to the form (\ref{Lambdaconf}). This implies that we can always reproduce the solutions of Einstein equations plus a cosmological constant with $\nabla_\mu A^\mu$ playing the role of the effective cosmological constant.

\section{Quantum fluctuations during inflation}

Let us consider an explicit example which is given by the
quantization in an inflationary de Sitter spacetime with
$a(\eta)=-1/(H_I\eta)$, with $H_I$ the constant Hubble parameter
during inflation. The explicit solution in the case $\xi=1/3$ for
the normalized scalar state is:
\begin{align}
\A_{0k}^{(s)}=&\frac{-1}{(2\pi)^{3/2}}\frac{i}{\sqrt{2k}}\left\{k\eta
e^{-ik\eta}+\frac{1}{k\eta}\left[\frac{1}{2}(1+ ik\eta)
e^{-ik\eta}-k^2\eta^2e^{ik\eta}E_1(2ik\eta)\right]\right\}e^{i\vec
k \vec x} \nonumber\\\nonumber\\
\A_{\parallel
k}^{(s)}=&\frac{1}{(2\pi)^{3/2}}\frac{1}{\sqrt{2k}}\left\{(1+ik\eta)e^{-ik\eta}
-\left[\frac{3}{2}e^{-ik\eta}+(1-ik\eta)e^{ik\eta}
E_1(2ik\eta)\right]\right\}e^{i\vec k \vec
x}\nonumber\\\label{scalar}
\end{align}
where $E_1(x)=\int_1^\infty e^{-tx}/tdt$ is the exponential
integral function. Using this solution, we  find:
\begin{eqnarray}
\nabla_\mu\A^{(s)\mu}_k=-\frac{a^{-2}(\eta)}{(2\pi)^{3/2}}\frac{ik}{\sqrt{2k}}\frac{3}{2}\frac{(1+ik\eta)}
{k^2\eta^2}e^{-ik\eta+i\vec k \vec x}
\end{eqnarray}
so that the field is suppressed in the sub-Hubble limit as
$\nabla_\mu\A^{(s)\mu}_k\sim \Od((k\eta)^{-2})$. 

On the other hand, from the energy density given by $\rho_A=
T^0_{\;\;0}$, we obtain in the sub-Hubble limit the corresponding
Hamiltonian, which is given by:
\begin{eqnarray}
H=\frac{1}{2}\int \frac{d^3\vec
k}{a^4(\eta)}k\sum_{\lambda=1,2,s}\left[ {\bf
a}_{\lambda}^{\dagger}(\vec k){\bf a}_{\lambda}(\vec k) +{\bf
a}_{\lambda}(\vec k){\bf a}_{\lambda}^{\dagger}(\vec k) \right].
\end{eqnarray}
We see that the theory does not contain negative energy states
(ghosts).

Also, from (\ref{scalar}) it is possible to obtain the dispersion
of the effective cosmological constant during inflation:
\begin{eqnarray}
\langle 0\vert(\nabla_\mu\A^{\mu})^2\vert 0
\rangle=\int\frac{dk}{k}P_A(k)
\end{eqnarray}
with $P_A(k)=4\pi k^3\vert\nabla_\mu\A^{(s)\mu}_k\vert^2 $. In the
super-Hubble limit, we get  in a 
 quasi-de Sitter inflationary phase characterized by a slow-roll
parameter $\epsilon$:
\begin{eqnarray}
P_{\nabla A}(k)=\frac{9H_{k_0}^4}{16\pi^2}
\left(\frac{k}{k_0}\right)^{-4\epsilon}
\label{PE}
\end{eqnarray}
where $H_{k_0}$ is the Hubble parameter when the 
$k_0$ mode left the horizon \cite{EM11,EM12}. Notice that this
result implies that $\rho_A\sim (H_{k_0})^4$. The measured value of
the cosmological constant then requires $H_{k_0}\sim 10^{-3}$ eV,
which corresponds to an inflationary scale  $M_I\sim 1$ TeV.
Thus we see that the cosmological constant scale can be naturally
explained in terms of physics at the electroweak scale.
This is one of the most relevant aspects of the present model
in which, unlike existing dark energy theories based on scalar fields, 
dark energy can be generated without including any potential term
or dimensional constant. It is also interesting to note that the required scale is the electroweak scale, since for higher energies electromagnetism becomes unified with the electroweak interactions so that it would not make sense to speak about photons anymore.

On the other hand, despite the fact that the background evolution in the present case
is the same as
in $\Lambda$CDM, the evolution of metric perturbations could
be different. We have calculated the evolution of metric,
matter density and
electromagnetic perturbations \cite{EM3}. The propagation speeds
of scalar, vector and tensor perturbations are found
to be real and equal to the speed of light, so that the theory is
classically stable.
On the other hand, it is
possible to see that all the parametrized post-Newtonian (PPN) parameters
\cite{Will}
agree with those of General Relativity,  i.e. the theory is compatible
with all the local gravity constraints for any value
of the homogeneous background electromagnetic field \cite{EM11,EM12,viability}.

Concerning the evolution of
scalar perturbations, 
we find
that  the only relevant deviations with respect to $\Lambda$CDM
appear on large scales $k\sim H_0$ and that
they depend on the primordial
spectrum of electromagnetic fluctuations. However,
the effects on the CMB temperature and matter power spectra 
are compatible with observations except for very large primordial
fluctuations \cite{EM3}.

\section{Generation of cosmic magnetic fields}

By looking at the equations of motion, we realize that the $\xi$-term 
can be interpreted as a conserved effective current: $-\xi\nabla^\mu(\nabla_\nu A^\nu)\equiv J_{\nabla\cdot A}^\mu$
which, according to (\ref{minimal}), satisfies the conservation
equation $\nabla_\mu J_{\nabla\cdot A}^\mu=0$. More precisely, the equations of motion can be recast in the form:
\begin{eqnarray}
\nabla_\nu F^{\mu\nu}=J^\mu_T
\end{eqnarray}
with $J^\mu_T=J^\mu+J^\mu_{\nabla\cdot A}$ and $\nabla_\mu
J^\mu_T=0$. Physically, this means that, while the new scalar mode
can only be excited gravitationally, once it is
produced it will generally give rise to an effective source of electromagnetic
fields. Therefore,  the modified theory is described by 
ordinary Maxwell's equations with an additional "external" current.
For an observer with four-velocity
$u^\mu$ moving with the cosmic plasma, 
it is possible to  decompose the Faraday tensor in its 
electric and magnetic parts  
as: $F_{\mu\nu}=2E_{[\mu}u_{\nu ]}+\frac{\epsilon_{\mu\nu\rho\sigma}}
{\sqrt{g}}B^\rho u^\sigma$, where $E^\mu=F^{\mu\nu} u_\nu$ and
 $B^\mu=\epsilon^{\mu\nu\rho\sigma} /(2\sqrt{g})F_{\rho\sigma}u_\nu$.
Due to the infinite conductivity of the plasma, 
Ohm's law $J^\mu-u^\mu u_\nu J^\nu=\sigma F^{\mu\nu} u_\nu$
implies  $E^\mu=0$.  
Therefore, in that case the only contribution  would come
from  the magnetic part.  
Thus, from Maxwell's equations, we can get:
\begin{eqnarray}
F^{\mu\nu}_{\;\;\;\; ;\nu}u_{\mu}=
\frac{\epsilon^{\mu\nu\rho\sigma}}
{\sqrt{g}}B_\rho u_{\sigma \,;\nu}u_\mu=J^{\mu}_{\nabla\cdot A}u_{\mu}
\end{eqnarray}
where we have made use of the electric neutrality of the cosmic
plasma, i.e. $J_\mu u^\mu=0$. For  comoving observers in a FLRW metric we can write
(see also Ref. \refcite{FC}):
\begin{eqnarray}
\vec \omega\cdot \vec B=\rho_g^0
\label{mag}
\end{eqnarray}
where $\vec v=d\vec x/d\eta$ is the conformal
time fluid velocity, $\vec\omega=\vec \nabla\times \vec v$ is 
the fluid 
vorticity, $\rho_g^0$ is the effective charge density today   
and the $\vec B$ components scale as $B_i\propto 1/a$ as can be
easily obtained from $\nabla_\mu \tilde{F}^{\mu\nu}=0$, with $\tilde{F}$ the dual of the Faraday 
tensor,
to the lowest order in $v$. Therefore, because of the presence of the effective charge density generated by the scalar state, we shall have both magnetic field and vorticity. Moreover, from the cosmological evolution of the magnetic field and the electric charge density given above, we obtain that  $\vert\vec\omega\vert
\propto a$, from radiation era until present.

We should remind here that $\nabla_\nu A^\nu$ is constant on super-Hubble scales
 and starts
decaying as $1/a$ once the mode reenters the Hubble radius. Thus, today, 
a mode $k$ will have been suppressed by a 
factor  $a_{in}(k)$ (we are assuming that the 
scale factor today is $a_0=1$). This factor will be given by: 
$a_{in}(k)=\Omega_M H_0^2/k^2$ for 
modes entering the Hubble radius in the matter era, i.e. for 
$k<k_{eq}$ with $k_{eq}\simeq (14\, \mbox{Mpc})^{-1}\Omega_Mh^2$ 
the value of the mode which entered
at matter-radiation equality. For $k>k_{eq}$ we have 
$a_{in}(k)=\sqrt{2\Omega_M}(1+z_{eq})^{-1/2}H_0/k$.
It is then possible
to compute from (\ref{PE}) 
the corresponding power spectrum for the effective 
electric charge
density today $\rho_g^0=J_{\nabla\cdot A}^0=-\xi\partial_0(\nabla_\nu A^\nu)$.
Thus from:
\begin{eqnarray} 
\langle\rho(\vec k)\rho^*(\vec h)\rangle=
(2\pi)^3\delta(\vec k-\vec h)\rho^2(k)
\end{eqnarray}
we define 
$P_\rho(k)=\frac{k^3}{2\pi^2}\rho^2(k)$, which is given by:
\begin{eqnarray}
P_\rho(k)=\left\{
\begin{array}{cc}
0, & k<H_0\\
& \\
\frac{\Omega_M^2H_0^2 H_{k0}^4}{16\pi^2}
\left(\frac{k}{k_0}\right)^{-4\epsilon-2},& H_0<k<k_{eq}\\
&\\
\frac{2\Omega_M H_0^2 H_{k0}^4}{16\pi^2(1+z_{eq})}
\left(\frac{k}{k_0}\right)^{-4\epsilon},&  k>k_{eq}.
\end{array}\right. 
\label{chPE}
\end{eqnarray}
Therefore the corresponding charge variance will read:
 $\langle\rho^2\rangle=\int \frac{dk}{k}P_\rho(k)$.
For modes entering the Hubble radius in the
radiation era, the power spectrum is nearly scale
invariant. Also, due to the constancy of $\nabla_\nu A^\nu$ on super-Hubble 
scales, 
the effective charge density power spectrum is negligible on such scales, so that
we do not expect magnetic field nor vorticity generation on those
scales. Notice that, on sub-Hubble scales, the effective charge density
generates longitudinal electric fields 
whose present amplitude would be precisely $E_L\simeq \nabla_\nu A^\nu$.
This implies that a generic prediction of the extended theory would be 
the existence of a cosmic background of longitudinal electric waves, whose 
power spectrum is given by (\ref{PE}).

\begin{figure}[t]
\begin{center}
{\epsfxsize=6.25cm\epsfbox{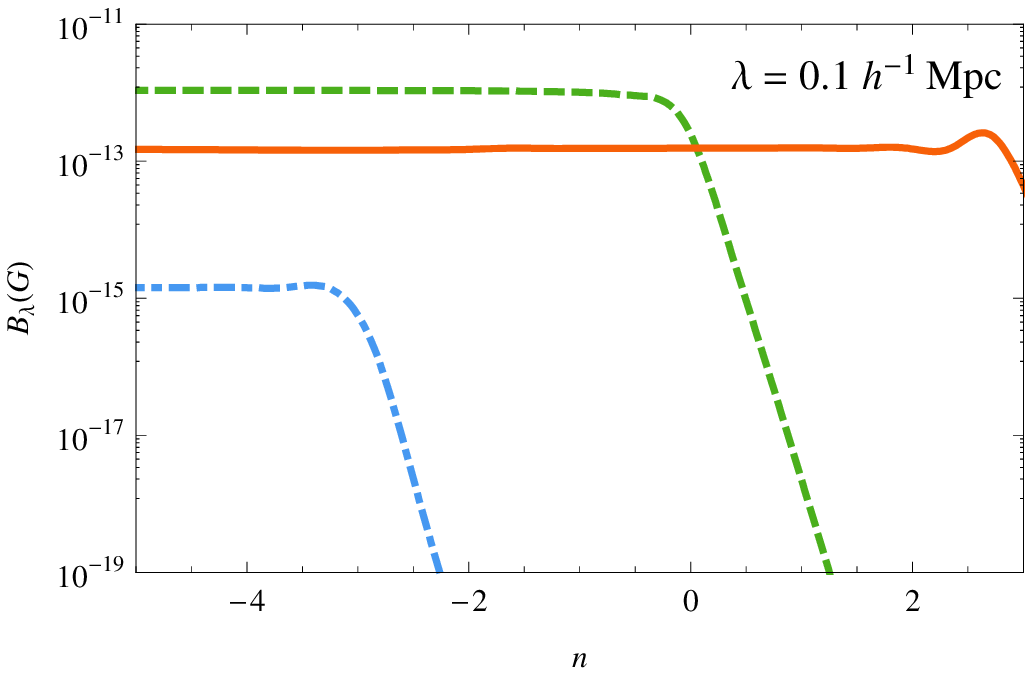}}
{\epsfxsize=6.25cm\epsfbox{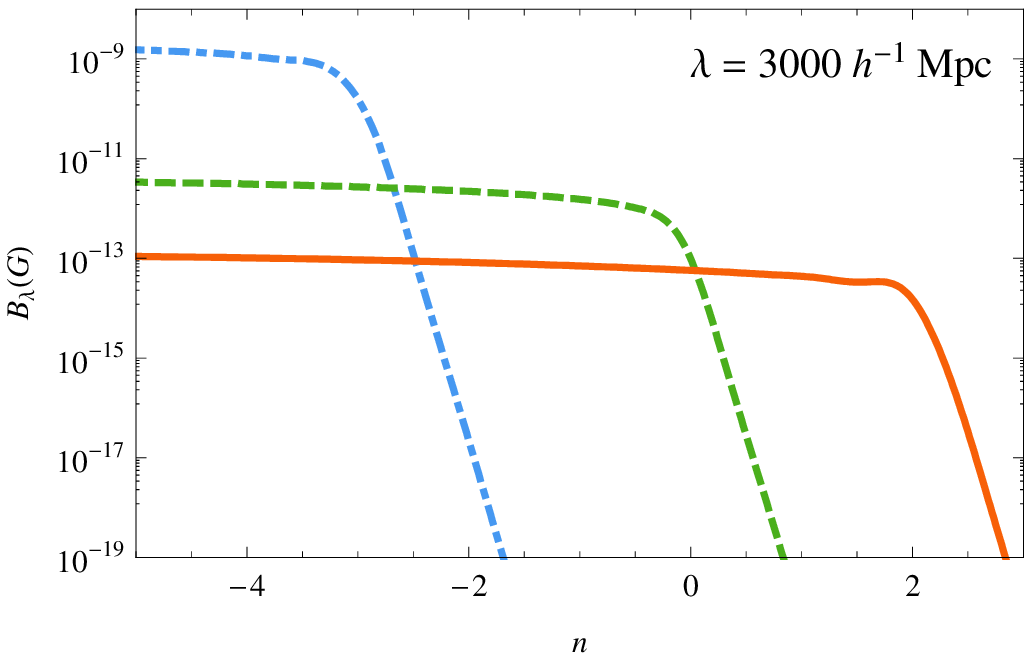}} 
\vspace*{-0.3cm}
\caption{ Lower limits on the magnetic
fields generated  on galactic scales (left panel)
and Hubble horizon scales (right panel) in terms
of the magnetic spectral index $n$ for different values
of the vorticity spectral index $m$. Dot-dashed blue for $m=0$, dashed 
green
for $m\simeq -3$ and full red for $m\simeq -5$.}
\end{center}
\end{figure} 
 Using (\ref{mag}), it is possible to translate the existing
upper limits on vorticity coming from CMB anisotropies \cite{FC} into  
{\it lower} limits on the amplitude of the magnetic fields 
generated by this mechanism. This indeed represents a discriminating signature of this model because the presence of such magnetics fields is a clear prediction of the theory. We will consider for simplicity
magnetic field and vorticity as gaussian stochastic variables
such that:
\begin{eqnarray}
\langle B_i(\vec k)B_j^*(\vec h)\rangle&=&
\frac{(2\pi)^3}{2}P_{ij}\delta(\vec k-\vec h)B^2(k)\nonumber \\ 
\langle\omega_i(\vec k)\omega_j^*(\vec h)\rangle&=&
\frac{(2\pi)^3}{2}P_{ij}\delta(\vec k-\vec h)\omega^2(k)
\end{eqnarray}
with $B^2(k)=B k^n$, $\omega^2(k)=\Omega k^m$
and where $P_{ij}=\delta_{ij}-\hat k_i\hat k_j$ is 
introduced because of the 
transversality properties of $B_i$ and $\omega_i$.
The spectral indices $n$ and $m$ are in principle arbitrary. 
In Fig. 2 we show  the 
lower limits on the magnetic fields
generated by this mechanism on  scales $\lambda=0.1 h^{-1}$ Mpc,
and  $\lambda=3000 h^{-1}$ Mpc, also for inflation at the
electroweak scale.     
 We see that fields can be generated with sufficiently 
large amplitudes in order to seed a galactic
dynamo or even to account for observations  just by collapse
and differential rotation of the protogalactic cloud \cite{magnetic1}.
Moreover, they could be also compatible with recent extra-galactic 
observations\cite{extragalactic1,extragalactic2}.

\section{Non-minimal couplings}

We shall end by generalizing the previous results to include non-minimal couplings to curvature:
\begin{eqnarray}
S=\int d^4x\sqrt{-g}\left[-\frac{1}{4}F_{\mu\nu}F^{\mu\nu}
+\frac{\xi}{2}(\nabla_\mu A^\mu)^2
+\sigma G_{\mu\nu}A^\mu A^\nu\right].
\label{action}
\end{eqnarray}
Notice that we are not introducing any new dimensional parameter, being
$\sigma$ a dimensionless constant. The particular non-minimal coupling has been chosen so that the induced effective electromagnetic current $J_{g}^\mu=2\sigma G^{\mu \nu}A_\nu$ is covariantly conserved in the weak field limit, which is achieved thanks to the divergenceless of the Einstein tensor $G_{\mu\nu}$. The most stringent constraint on $\sigma$ comes from PPN measurements and it is given by $\vert\sigma\vert \lsim 10^{-5}$. Moreover, the smallness of $\sigma$ guarantees the stability of the theory and that the cosmological evolution of the homogeneous mode 
becomes modified in a negligible way by the
 presence of the non-minimal coupling. 
This ensures that the  
inflationary generation and cosmological evolution discussed in previous
sections for the minimal theory is also a 
good description in the non-minimal case 
  (see Ref. \refcite{nonminimal} for a more detail discussion).

In the following we shall show how the non-minmal coupling gives rise to non-trivial electromagnetic effects generated by massive bodies even if they are electrically neutral. To that end, we shall assume a small perturbation around Minkowski with a constant background vector field of the form\footnote{The background vector field is supposed to be given by the cosmological one which only varies on cosmological timescales and whose present value would be $\bar A_0 \simeq 0.3\,M_p$ according to the observed
dark energy density.} $A_\mu=\bar{A}_0\delta ^0_\mu$. Thus, the effective current becomes $J_{g}^\mu=2\sigma G^{\mu 0}\bar{A}_0$ and, 
using 
Einstein equations to relate $G^{\mu\nu}$ to the matter content, we 
obtain:
\begin{eqnarray}
J_{g}^\mu=16\pi G\sigma T^{\mu 0}\bar A_0
\end{eqnarray}
so that the effective 
electromagnetic current is essentially determined by the four-momentum 
density.  Moreover, if we assume $T^{\mu\nu}=
(\rho+p)u^\mu u^\nu-p\eta^{\mu\nu}$ at first order, 
we can see that the energy density
of any perfect fluid 
has an associated electric charge density given, for small velocities, 
by: 
\begin{equation}
\rho_g=J_g^0=16\pi G\sigma\rho \bar A_0\label{effectiveq}
\end{equation}
and the three-momentum density generates an electric current density 
given by 
\begin{equation}
\vec{J}_g=16\pi G\sigma (\rho+p)\vec{v}\bar A_0
\end{equation}
Thus, this theory effectively realizes the old conjecture by Schuster, Einstein  
and Blackett \cite{old1,old2,old3} 
of gravitational magnetism, i.e. neutral mass currents can generate
electromagnetic fields. 

In the case of a particle of mass $m$ at rest, 
(\ref{effectiveq}) introduces a small contribution to the 
{\it active} electric charge 
(the source of the electromagnetic field), given by
$\Delta q= 16\pi G\sigma m \bar A_0\simeq 15\sigma (m/M_P)$, 
but leaving unmodified its
{\it passive} electric charge (that determining the coupling 
to the electromagnetic field).  This effect implies a difference in the active charge of electrons and protons because of their mass difference and, furthermore, provides neutrons with a non-vanishing active electric charge. However, the effect is 
very small in both cases $\Delta q\simeq 4\sigma  10^{-18}e$
where  $e=0.303$ is the 
electron charge 
in Heaviside-Lorentz units. Present limits on the 
electron-proton charge asymmetry  and neutron charge are both of the order
$10^{-21}e$ \cite{limits}, 
implying $\vert \sigma\vert\lsim 10^{-3}$ which is less stringent than
the PPN limit discussed before.

On the other hand, for any compact object, the effective electric current will  
generate an intrinsic magnetic moment  given by:
\begin{equation}
\vec{m}=\beta\frac{\sqrt{G}}{2}\vec{L}\label{S-Blaw}
\end{equation}
with $\vec L$ the corresponding angular momentum and 
$\beta=16\pi \sqrt{G}\sigma \bar A_0$ a constant parameter.
This relation resembles the 
Schuster-Blackett law, which is a purely empirical relation between
 the magnetic moment  and the angular momenta found in a wide
range of astrophysical objects from planets, to galaxies, including those
related to the presence of rotating neutron stars such as GRB 
or magnetars \cite{opher1,opher2}. In any case,  
the observational evidence on this relation  
is still not conclusive. From observations, 
the $\beta$ parameter ranges from 0.001 to 0.1. Imposing the PPN limits on the $\sigma$ parameter, we find
$\beta\lsim 10^{-4}$, which is just below the observed range.
Thus for a typical spiral galaxy, a direct calculation provides:
$B\sim \sigma 10^{-4}$ G, i.e. according to the PPN limits, the field
strength could reach $10^{-9}$ G without amplification.

\section{Discussion}

We have reviewed the extended electromagnetic theory in which a gauge-fixing term is promoted into a physical contribution  in the fundamental action. The quantization of the free theory can be performed without having to impose any subsidiary condition at the price of introducing an additional degree of freedom. This new mode can be gravitationally produced from quantum fluctuations during inflation and its amplitude will be determined precisely by the scale at which inflation takes place.  In the subsequent cosmological evolution, this new mode has two effects. On super-Hubble scales, it behaves as an effective cosmological constant whose observed value can be explained if inflation occurred at the electroweak scale. On sub-Hubble scales, the additional mode gives rise to a stochastic background  of longitudinal electric waves that generate magnetic fields from sub-galactic scales up to the 
present Hubble radius. On the other hand, we have also considered the effects of non-minimal 
couplings in the presence of the temporal electromagnetic mode and shown that in such a case, 
neutral massive objects can act as sources of electromagnetic fields thus implementing the 
old conjecture by Schuster, Einstein and Blackett of gravitational magnetism.  

The theory studied in this work shows how a modification of  electromagnetism
which does not require the introduction of new fields, dimensional parameters
or potential terms  could provide a simple explanation for
the tiny value of the cosmological constant and, at the same time, a
mechanism for the generation of magnetic fields on cosmological scales.
Some open questions still remain to be studied such as the inclusion of electromagnetic 
interactions in the theory, since all the analysis performed so far are limited to the free theory. Also,  
it would be interesting to know the behaviour of the new mode in more general background space-times 
in order to determine the viability of the model from the theoretical and phenomenological
point of views. Finally,  the possibility of detecting  the longitudinal electric wave background generated during inflation could provide a clear signal of the modification of electromagnetism
on cosmological scales.

\vspace{0.2cm}

{\em Acknowledgments:}
 This work has been  supported by
MICINN (Spain) project numbers
FIS 2008-01323 and FPA
2008-00592, CAM/UCM 910309 and MICINN Consolider-Ingenio 
MULTIDARK CSD2009-00064. JBJ is also supported by the Ministerio de Educaci\'on under the postdoctoral contract EX2009-0305.
\vspace{0.2cm}


\begin{thebibliography}{99}
\bibitem{limit} A.~S.~Goldhaber, M.~M.~Nieto,
  Rev.\ Mod.\ Phys.\  {\bf 82 } (2010)  939-979.
\bibitem{galactic1}  L.~M.~Widrow,
  Rev.\ Mod.\ Phys.\  {\bf 74} (2002) 775.
 \bibitem{galactic2} R.~M.~Kulsrud and E.~G.~Zweibel,
  Rept.\ Prog.\ Phys.\  {\bf 71} (2008) 0046091;.
  \bibitem{galactic3}P.~P.~Kronberg,
  Rept.\ Prog.\ Phys.\  {\bf 57} (1994) 325.
\bibitem{extragalactic1} A. Neronov and I. Vovk, 
Science {\bf 328} (2010) 73; 
\bibitem{extragalactic2}F.~Tavecchio, et al.,  
MNRAS, 406: L70-L74,2010.
\bibitem{extragalactic3} S.~i.~Ando, A.~Kusenko, Astrophys.\ J.\  {\bf 722 } (2010)  L39.
 \bibitem{extragalactic4}A.~Neronov, D.~V.~Semikoz, P.~G.~Tinyakov {\it et al.}, 
[arXiv:1006.0164 [astro-ph.HE]].
\bibitem{Parker} A.~Higuchi, L.~Parker and Y.~Wang,
  Phys.\ Rev.\  D {\bf 42}, 4078 (1990).
\bibitem{EM2} J. Beltr\'an Jim\'enez and A.L. Maroto, 
Phys.\ Lett.\  B {\bf 686} (2010) 175.
\bibitem{ghosts1}
A.~R.~Zhitnitsky,
  Phys.\ Rev.\  D {\bf 82} (2010) 103520.
\bibitem{ghosts2}  N.~Ohta, Phys. Lett. {\bf B695}: 41-44 (2011).

\bibitem{Birrell}
  Birrell, N.D. and Davies, P.C.W.,
  {\it Quantum fields in curved space},
  Cambridge University Press,
  1982,
  ISBN: 0-521-27858-9.

\bibitem{Adler:1976jx}
S.~L. Adler, J.~Lieberman, and Y.~J. Ng,   {\em Ann. Phys.} {\bf 106} (1977) 279.

\bibitem{Brown:1986tj}
M.~R. Brown and A.~C. Ottewill,,  {\em
  Phys. Rev.} {\bf D34} (1986) 1776--1786.

\bibitem{Pfenning:2001wx}
M.~J. Pfenning,
  {\em Phys. Rev.} {\bf D65} (2002) 024009.

\bibitem{EM11} J. Beltr\'an Jim\'enez and A.L. Maroto, 
JCAP {\bf 0903} (2009) 016. 
\bibitem{EM12} J. Beltr\'an Jim\'enez and A.L. Maroto, Int.\ J.\ Mod.\ Phys.\  D {\bf 18} (2009) 2243.

\bibitem{EM3}
J.~B.~Jimenez, T.~S.~Koivisto, A.~L.~Maroto and D.~F.~Mota,
 JCAP {\bf 0910} (2009) 029.
 
 \bibitem{Will} C. Will, {\it Theory and experiment in gravitational physics},
Cambridge University Press, (1993).
\bibitem{viability}J.~B.~Jimenez and A.~L.~Maroto,  JCAP {\bf 0902} (2009) 025.

\bibitem{FC} C.~Caprini and P.~G.~Ferreira,
  JCAP {\bf 0502} (2005) 006.


\bibitem{magnetic1}  J. Beltr\'an Jim\'enez, A.~L.~Maroto, {\em Phys. Rev.} {\bf D83}: 023514 (2011). 

\bibitem{nonminimal} J. Beltr\'an Jim\'enez, A.~L.~Maroto,
 JCAP 1012: 025 (2010). 

\bibitem{old1} A. Schuster,   Proc. Lond. Phys. Soc. {\bf 24} (1912) 121.
\bibitem{old2} A. Einstein,  Schw. Naturf. Ges. Verh. 105 Pt. 2, 85 (1924) 
S. Saunders S and H.R. Brown,  Philosophy of Vacuum
(Oxford: Clarendon) (1991).
\bibitem{old3}P. M. S. Blackett,   Nature {\bf 159} (1947) 658.

\bibitem{limits} C. Amsler et al. (Particle Data Group), 
Phys.\ Lett.\  B {\bf 667} (2008) 1.
\bibitem{opher1}R. Opher and U.~F.~Wichoski,
  Phys.\ Rev.\ Lett.\  {\bf 78} (1997) 787.
\bibitem{opher2} R. da Silva de Souza, R. Opher, JCAP {\bf 1002} (2010) 022.



\end{thebibliography}

\end{document}